\documentclass[conference]{IEEEtran}
\IEEEoverridecommandlockouts
\usepackage{cite}
\usepackage{amsmath,amssymb,amsfonts}
\usepackage{graphicx}
\usepackage{textcomp}
\usepackage[dvipsnames]{xcolor}
\usepackage{booktabs}
\usepackage{multicol}
\usepackage{multirow}
\usepackage{url}
\usepackage[linesnumbered,ruled,vlined]{algorithm2e}

\SetCommentSty{mycommfont}
\SetKwInOut{Input}{Input}
\SetKwInOut{Output}{Output}

\definecolor{LSC}{rgb}{1,0.07,0.0}

\usepackage{fancyhdr}
\fancypagestyle{firstpage}{
\fancyhf{}
\fancyhead[C]{To appear at the 2022 IEEE Computer Society Annual Symposium on VLSI
 – ISVLSI 2022.} 
}

\begin{document}

\title{Optimization of BDD-based\\
Approximation Error Metrics Calculations\vspace*{-0.5em}}

\author{
\IEEEauthorblockN{Vojtech Mrazek}
\IEEEauthorblockA{\textit{Brno University of Technology, Faculty of Information Technology} \\
Brno, Czech Republic \\
mrazek@fit.vutbr.cz
\vspace*{-1em}}
}

\newcommand{\repo}{\textcolor{blue}{\url{https://github.com/ehw-fit/bdd-evaluation}}}

\newcommand{\subtract}[1]{\mathrm{subtract}(#1)}
\newcommand{\add}[1]{\mathrm{add}(#1)}
\newcommand{\sat}[1]{\mathrm{SAT}(#1)}
\newcommand{\satp}[1]{\mathrm{\#SAT}(#1)}
\newcommand{\sign}[1]{#1_{\mathrm{sign}}}

\maketitle

\thispagestyle{firstpage}

\begin{abstract}
Software methods introduced for automated design of approximate implementations of arithmetic circuits rely on fast and accurate evaluation of approximate candidate implementations. To accelerate the evaluation of circuit error, we propose four novel algorithms for the exact worst-case and mean absolute error analysis based on Binary Decision Diagrams. As these algorithms do not compute any absolute values in the characteristic function, which basically compares a candidate approximate circuit with a golden circuit, the error evaluation is significantly faster than the standard BDD-based error analysis. On average, the proposed algorithms are three times faster (in some cases, 30 times faster) than the baseline for 8- to 32-bit approximate adders. These results were obtained from more than 49 thousand runs with different configurations of the method.
The proposed error evaluation algorithms are available as an open-source software~\repo.

\end{abstract}


\section{Introduction}

Minimizing power consumption is one of the most significant objectives for current digital circuit design. In many applications, such as image or signal processing, the users are also often willing to accept certain errors if they are accompanied by reduced power consumption or other improvements. This paradigm, called approximate computing~\cite{jieOrshanskyAXcomp}, systematically tries to find the best trade-offs between the quality of output and hardware parameters such as area, power consumption, or latency. One of the most widely used methods is replacing the standard arithmetic circuits with approximate implementations~\cite{jieOrshanskyAXcomp}. The circuit approximation techniques can be classified as manual (e.g.,~\cite{jiang}) and automated (e.g.,~\cite{hashemi2018blasys,ceska:iccad17}). From a broader perspective, the automated methods usually provide better trade-offs between the error and other circuit parameters. These methods typically start with an exact circuit and modify it using the following iterative scheme: (i) generate candidate approximate circuits, (ii) evaluate these circuits, and (iii) select the best candidate for the next iteration. As many candidate circuits are usually generated, it is essential to quickly analyze their error~\cite{ceska:iccad17}.

Let $f: \mathbb{B}^n \rightarrow  \mathbb{B}^m$ be an $n$-input $m$-output Boolean function that describes the correct functionality (the accurate function) and $f': \mathbb{B}^n \rightarrow  \mathbb{B}^m$  be an approximation of it with error $\varepsilon$, formally $\forall x \in \mathbb{B}^n: f(x) \approx f'(x); f'(x) = f(x) + \varepsilon(x)$. If there is no analytical model of the error function $\varepsilon$ (which can, in fact, be easily derived only for a few approximation techniques such as \textit{bit-truncation}), the error of $f'$ has to be determined for all possible input combinations, i.e. $|\mathbb{B}^n| = 2^n$. The evaluation time grows exponentially with $n$, but can effectively be reduced by parallel circuit simulation~\cite{hrbacek:gecco14} or formal verification techniques such as SAT, BDDs or algebraic solvers~\cite{ceska:iccad17,keszocze:access2022}.

The formal verification techniques achieve very good performance for the error evaluation~\cite{ceska:iccad17}. However, some of them do not scale for some types of circuits or error metrics. For example, SAT solving is applicable for determining the worst-case error but useless for the mean error. BDD-based methods are not applicable to structurally complex circuits such as multipliers.

In this paper, we deal with efficient error analysis methods using BDDs.
 The calculation is based on the so-called characteristic function $|f - f'|$ represented as a Boolean equation~\cite{soeken:bdd}. Keszocze~\cite{keszocze:access2022} introduced a technique that does not need to compute the absolute value which represents the performance bottleneck. On the other hand, this algorithm assumes that $\forall x: f' (x) \le f(x)$ and can produce only such solutions that satisfy this assumption. However, this assumption is not realistic is general-purpose approximation engines (such as~\cite{ceska:iccad17}) in which arbitrary candidate approximations must be evaluated.
 
  
This paper aims to propose new algorithms for the exact error analysis that  further simplify the characteristic function without postulating any specific constraint on the evaluated function $f'$.

We propose four error calculation algorithms (two for the worst-case error and two for the mean-absolute error). These algorithms omit the absolute value calculation while the average error evaluation time is reduced $3\times$ compared to the standard BDD-based approach. Moreover, they also reduce the memory complexity, and in some cases, the speedup is $10-30\times$.



The proposed error analysis algorithms were employed in a circuit approximation method based on CGP, which is inspired in~\cite{ceska:iccad17}.
In our experiments, the error evaluation was executed for billions of adders.
The resulting approximate adders exhibit better trade-offs than the approximate adders published in the EvoApprox library~\cite{mrazek:date16lib}. All approximate implementations are available for download (TBD). The C++ implementation of the error analysis algorithms is provided as an open-source software \repo.



\section{Approximate computing}
Approximate computing aims to introduce an error into a software or hardware implementation. It can be done on various levels: algorithmic, functional, or physical. We focus on the functional approximation (i.e., modification of Boolean functions describing the logic behavior of digital circuits) of arithmetic circuits that may be used as basic building blocks in complex systems.

\subsection{Error metrics}

There are various application-independent error metrics used in the literature~\cite{jiang}. The \emph{worst-case arithmetic error}, sometimes denoted as \emph{error magnitude} or \emph{error significance}, is defined as
\begin{equation}
e_{wce}(f, f’) = \max_{\forall x \in \mathbb{B}^n} |int(f(x)) - int (f’(x))|,
\end{equation}
where $int(x)$ represents a function $int: \mathbb{B}^m \rightarrow \mathbb{Z}$ returning an integer value of the $m$-bit binary vector $x$. Typically, a natural unsigned binary representation is considered, i.e. $ int(x) = \sum_{i=1}^{m}{2^i \cdot x_i}$. The worst-case error represents the fundamental metric that is useful to guarantee that the approximate output differs from the correct output by at most the error bound $e$.

The \emph{average-case arithmetic error} (also known as \emph{mean absolute error}) is defined as the sum of absolute differences in magnitude between the original and approximate circuit, averaged over all inputs:
\begin{equation}
e_{mae}(f, f’) = 2^{-n} \sum_{\forall x \in \mathbb{B}^n} {| int(f(x)) - int(f’(x))| }.
\end{equation}



Note that the values produced by the ``absolute'' error metrics $e_{mae}$ and $e_{wce}$ can be very high. Hence, these values can be expressed relatively to the output range, using division by $2^{m} - 1$, i.e. the maximal output value. 

In the metric error formulas, the enumeration of all possible input vectors is present. For $n$ greater than approx. 24, it is not feasible to enumerate $\mathbb{B}^n$ in a reasonable time even when a highly parallel circuit simulator is employed. 

This issue can partly be eliminated by utilizing a formal verification approach for the error calculation. Formal techniques typically construct a virtual miter circuit to compute the characteristic function. Reduced Ordered Binary Decision Diagrams (BDD) or SAT solving-based methods were employed in the area of approximate circuits. ROBDDs are suitable for the formal error analysis of less complex circuits (such as adders) for which the BDD does not grow exponentially with $n$. SAT solving can effectively handle determining the worst-case error for up to 32-bit multipliers but does not provide suitable results for other error metrics. Details of BDDs are elaborated in Section~\ref{sec:bdds}.

\subsection{Automated design techniques}

The automated approximation algorithms iteratively modify a given accurate circuit to achieve the best trade-off between the error and other circuit parameters. The most recent algorithms use various approaches for circuit representation and error evaluation. Soeken et al. proposed an algorithm that directly modifies the BDD~\cite{soeken:bdd,keszocze:access2022}. This algorithm satisfies the assumption of the fast error evaluation, but its design space is limited. Another approach is to use some internal circuit representation. Hashemi et al. divided the original circuit into subcircuits and approximated each of them. Finally, the error was evaluated for many candidate solutions~\cite{hashemi2018blasys}. A specific internal circuit representation is used in \textit{Cartesian Genetic Programming} (CGP)~\cite{ceska:iccad17} which is a heuristic design-space exploration algorithm. Candidate circuits are represented as directed acyclic graphs in a fixed grid of nodes. The algorithm randomly changes the connections between nodes and the nodes' functions. It uses the following fitness function for a given error metric $e$ (e.g., $e_{mae}$ or $e_{wce}$) and the acceptable error threshold $\tau$
\begin{equation}\label{eq:fitness}
\mathbf{F}(f') = 
\begin{cases}
size(f') & \text{if } e(|f - f'|) \leq \tau \\
\infty & \text{otherwise} \\
\end{cases}.
\end{equation}
This algorithm starts with an accurate solution and reduces its size while keeping the error constraint satisfied. The result of the search is the smallest circuit satisfying the constraint $\tau$.



\subsection{BDDs}
\label{sec:bdds}

This work exploits the Reduced Ordered Binary Decision Diagrams (ROBDDs, also referred to as BDDs). ROBDD is a canonical rooted acyclic graph-based representation of a Boolean function, where each \textbf{node} represents either an input variable or terminal node, and the edges represent the value assignment. There are two terminal nodes -- \textit{true} and \textit{false}. ROBDDs have been traditionally used to solve the equivalence checking problem due to their canonical property (i.e., the ROBDDs of two logic functions are isomorphic if the functions are functionally equivalent and the input variables have the same order). The tools, developed for constructing and manipulating ROBDDs, can exactly answer the following questions: (1) whether a variable (signal) in the circuit is satisfiable, i.e., whether an input vector exists which generates a true value of the signal; and (2) the probability that a selected variable is true. Although these questions are answered fast (by finding some paths in the ROBDD), the construction of the ROBDD is not trivial, and in some circuits such as multipliers, the ROBDDs grow exponentially with $n$.

Since there is no significant difference in the performance of major tools operating with ROBDDs~\cite{dijkBDD}, we selected BuDDy and CUDD tools, which also support dynamic variables reordering for evaluation. In these tools, the BDD node is represented by a structure. Starting with an input node, these tools allow adding new nodes by applying logical operations ($\neg, \wedge, \vee, \oplus, \dots$) to two existing nodes. Therefore, one can construct an arbitrary Boolean function.

\subsubsection{Conventions and notations}
Henceforth, we denote an $n$-input $m$-output Boolean function as $f$. Without loss of generality the functions can be represented by the BDDs $f=(f_n, f_{n-1}, \dots f_2, f_1)$ where each output bit $f_i$ is a BDD node. For integer functions, $\sign{f}$ denotes the sign bit $f_n$.

For one-output function (i.e., BDD node) the tools can provide the probability that a selected node $f_i$ is true
$$\satp{f_i} = 2^{-n}  \cdot |\{x : f_i(x) = \textrm{True}\}| $$
and a Boolean value $\sat{f_i}$ answering whether the node $f_i$ becomes \textit{true} for any input.

We also implemented two arithmetic Boolean functions as BDDs. The Addition function $\add{a, b}$ creates a Ripple-carry adder with output nodes of functions $a$ and $b$ as inputs. Similarly, subtraction $\subtract{a,b}$ is constructed in the same way, but the $b$ outputs are inverted, and the carry input is set. Note that the input vectors must be (sign-) extended to handle both signed and unsigned variants of $a$ and $b$.

\subsubsection{Existing algorithms (\textbf{BASELINE})}
In \cite{soeken:bdd}, Soeken et al. proposed the error analysis based on calculating characteristic function $r = |f(x) - f'(x)|$. For the resulting vector $r$ they calculate the maximal (wce) and average (mae) values. 
Their best-performing algorithm is based on a binary search (Alg. \ref{alg:wce_baseline}). Firstly, the values are subtracted (ln. 1), which is captured by a signed function $\varepsilon$, and the absolute value is obtained (unsigned $r$). The absolute value has to deal with the two's complement and increments the value if the result is negative (ln. 4). The binary search starts with the MSB, computes the worst-case error $wce$ and also the function $\mu: \mu(x) = 1$ if and only if $r(x) = wce$. 

\begin{algorithm}[ht]
\caption{Baseline WCE calculation using two's complement}\label{alg:wce_baseline}
\Input{Functions $f$ and $f'$ as ROBDDs}
\Output{Worst-case error ($wce$)}
$\varepsilon \gets \subtract{f, f'}$  \tcp*{subtract results}

\For{$i \gets 1$ \KwTo $|\varepsilon| - 1$} {
    $r_i \gets \varepsilon_i \oplus \sign{\varepsilon} $ \tcp*{absolute value}
}
$r \gets \add{r, \{\sign{\varepsilon}\}}$  \tcp*{two's complement}

$ \mu \gets  $ True\;
$ wce \gets 0$\;
\For(\tcp*[h]{binary search}){$i \gets |r|$ \KwTo $1$ }{
    \If{$\sat{\mu \wedge r_i}$} {
        $wce \gets wce + 2^{i - 1}$\;
        $\mu \gets \mu \wedge r_i$\;
    }
}
\Return $wce$\;
\end{algorithm}

The average (mean absolute error) is calculated (Alg. \ref{alg:mae_baseline}) with the same characteristic function $r$. The weighted sum of positive-bit probabilities is computed. Finally, the sum is divided by $2^n$ while $n$ is the number of inputs of functions $f$ and $f'$.

\begin{algorithm}[ht]
\caption{Baseline MAE calculation using two's complement}\label{alg:mae_baseline}
\Input{Functions $f$ and $f'$ as ROBDDs}
\Output{Mean absolute error ($mae$)}
$\varepsilon \gets \subtract{f, f'}$  \tcp*{subtract results}

\For{$i \gets 1$ \KwTo $|\varepsilon| - 1$} {
    $r_i \gets \varepsilon_i \oplus \sign{\varepsilon} $ \tcp*{absolute value}
}

$r \gets \add{r, \{\sign{\varepsilon}\}}$  \tcp*{two's complement}

$ mae \gets 0$\;
\For{$i \gets 1$ \KwTo $|r|$} {
    $mae \gets mae + 2^{i-1} \cdot \satp{r_i}$ \;
}
\Return $mae$ \;
\end{algorithm}

Keszocze showed that the subtraction and absolute value calculation are the most time-consuming parts of the evaluation~\cite{keszocze:access2022},. He also proposed a fast algorithm for determining the MAE without the characteristic function. This algorithm needs to satisfy a condition that $\forall x: f'(x) \leq f(x)$. Automated iterative approximation algorithms typically cannot satisfy this condition, therefore, we are not able to employ this algorithm in a CGP-based approximation engine.

\section{Proposed MAE and WCE calculation}

We present improvements of the baseline error evaluation algorithms to omit the absolute value calculation. We propose two approaches --- one replaces the absolute value by approximation in ones' complement, the second one handles the positive and negative branches separately.

\subsection{Ones' complement \textbf{(ONES')}}
In the baseline Algorithms \ref{alg:wce_baseline},\ref{alg:mae_baseline} (ln. 4), the increment of $r$ vector by a sign bit is needless. In the proposed algorithm (Alg. \ref{alg:wce_ones}), we omitted this addition. We obtain the exact $wce$ for positive $r$; and ${wce}$ decreased by $1$ if the maximal error is in the negative $r$. Because this condition is expressed in the MSB node of $\varepsilon$, the error can be compensated (ln. 10 - 11).

\begin{algorithm}[ht]
\caption{Proposed WCE calculation using ones' complement}\label{alg:wce_ones}
\Input{Functions $f$ and $f'$ as ROBDDs}
\Output{Worst-case error (${wce}$)}
$\varepsilon \gets \subtract{f, f'}$  \tcp*{subtract results}

\For{$i \gets 1$ \KwTo $|\varepsilon| - 1$} {
    $r_i \gets \varepsilon_i \oplus \sign{\varepsilon} $ \tcp*{absolute value}
}

$ \mu \gets  $ True\; 
$ wce \gets 0$\;

\For(\tcp*[h]{binary search}){$i \gets |r|$ \KwTo $1$ }{
    \If{$\sat{\mu \wedge r_i}$} {
        $wce \gets wce + 2^{i - 1}$\;
        $ \mu \gets \mu \wedge r_i$\;
    }
}

\If(\tcp*[h]{complement corr.}){$\sat{\mu \wedge  \sign{\varepsilon}}$} {
    $wce \gets wce + 1$ \;
}
\Return $wce$\;
\end{algorithm}


A similar approach is used in the average error calculation (Alg. \ref{alg:mae_ones}). The error by 1 can be corrected by adding the proportion of negative numbers $\varepsilon$, i.e. $\satp{\sign{\varepsilon}}$ (ln. 7).

\begin{algorithm}[ht]
\caption{Proposed MAE calculation using ones' complement}\label{alg:mae_ones}
\Input{Functions $f$ and $f'$ as ROBDDs}
\Output{Mean absolute error ($mae$)}
$\varepsilon \gets \subtract{f, f'}$  \tcp*{subtract results}

\For{$i \gets 1$ \KwTo $|\varepsilon| - 1$} {
    $r_i \gets \varepsilon_i \oplus \sign{\varepsilon} $ \tcp*{absolute value}
}

$ mae \gets 0$\;
\For{$i \gets 1$ \KwTo $|r|$} {
    $mae \gets mae + 2^{i-1} \cdot \satp{r_i}$ \;
}

$mae \gets mae + \satp{\sign{\varepsilon}} $ \tcp*{correction}
\Return $mae$  \;
\end{algorithm}

\subsection{Omitting the absolute values \textbf{(NOABS)}}
The second proposed approach exploits the fact that we can build two BDD trees for the resulting vector $\varepsilon$ --- one for the positive part $\neg \sign{\varepsilon} \wedge \varepsilon_i$  and one for the negative part  $\sign{\varepsilon} \wedge \varepsilon_i$. The worst-case error calculation is shown in Alg. \ref{alg:wce_noabs}. Since the absolute error is calculated using XOR operations which make the resulting BDDs more complex, the WCE value is calculated directly from the signed result of subtraction $\varepsilon$. If the sign bit is negative, a binary search of the maximal value is performed as $wce_p$. On the other hand, the negative part is transformed to searching the minimal value $wce_n$. The subtraction result $\varepsilon$ is expressed as two's complement. Therefore, the negative $wce_n$ is incremented by 1, and the maximal value is returned (ln. 14).

\begin{algorithm}[ht]
\caption{Proposed WCE calculation without the absolute value (NOABS)}\label{alg:wce_noabs}
\Input{Functions $f$ and $f'$ as ROBDDs}
\Output{Worst-case error ($wce$)}
$\varepsilon \gets \subtract{f, f'}$  \tcp*{subtract results}

\tcc{positive error}
$ \mu_p \gets  \neg \sign{\varepsilon}$\;
$ wce_p \gets 0$\;
\For(\tcp*[h]{binary search}){$i \gets |\varepsilon| - 1$ \KwTo $1$ }{
    \If{$\sat{\mu_p \wedge \varepsilon_i}$} {
        $wce_p \gets wce_p + 2^{i - 1}$\;
        $\mu_p \gets \mu_p \wedge \varepsilon_i$\;
    }
}

\tcc{negative error}
$ \mu_n \gets  \sign{\varepsilon}$\;
$ wce_n \gets 0$\;
\For(\tcp*[h]{binary search}){$i \gets |\varepsilon| - 1$ \KwTo $1$ }{
    \If{$\sat{\mu_n \wedge \neg \epsilon_i}$} {
        $wce_n \gets wce_n + 2^{i - 1}$\;
        $\mu_n \gets \mu_n \wedge \neg \varepsilon_i$\;
    }
}
\Return $\max(wce_p, wce_n + 1)$\;
\end{algorithm}

The mean average error calculation (Alg. \ref{alg:mae_noabs}) employs two BDD trees --- for the positive part anded with the unset sign bit $\neg \sign{\varepsilon}$ and for the negative inverted anded with a set sign bit $\sign{\varepsilon}$. The negative part introduces an error by one, which can easily be compensated by adding the probability at line 6.

\begin{algorithm}[ht]
\caption{Proposed MAE calculation without the absolute value (NOABS)}\label{alg:mae_noabs}
\Input{Functions $f$ and $f'$ as ROBDDs}
\Output{Mean absolute error ($mae$)}
$\varepsilon \gets \subtract{f, f'}$  \tcp*{subtract results}
$ mae \gets 0$\;
\For{$i \gets 1$ \KwTo $|\varepsilon| - 1$} {
    $mae \gets mae + 2^{i-1} \cdot \satp{\varepsilon_i \wedge \neg \sign{\varepsilon}}$ \;
    $mae \gets mae + 2^{i-1} \cdot \satp{\neg \varepsilon_i \wedge  \sign{\varepsilon}}$ \;
}

$mae \gets mae + \satp{\sign{\varepsilon}}  $ \tcp*{complement corr.}
\Return $mae$ \;
\end{algorithm}

All the proposed algorithms follow the scheme of WCE and MAE calculation introduced by Soeken et al.~\cite{soeken:bdd} However, the absolute value calculation is partially (in ones') or totally (in noabs) transformed from the BDD function to mathematical corrections.

\section{Experiments}

\subsection{Experimental setup}
\begin{figure}[ht]
    \centering
    \includegraphics[width=0.8\columnwidth]{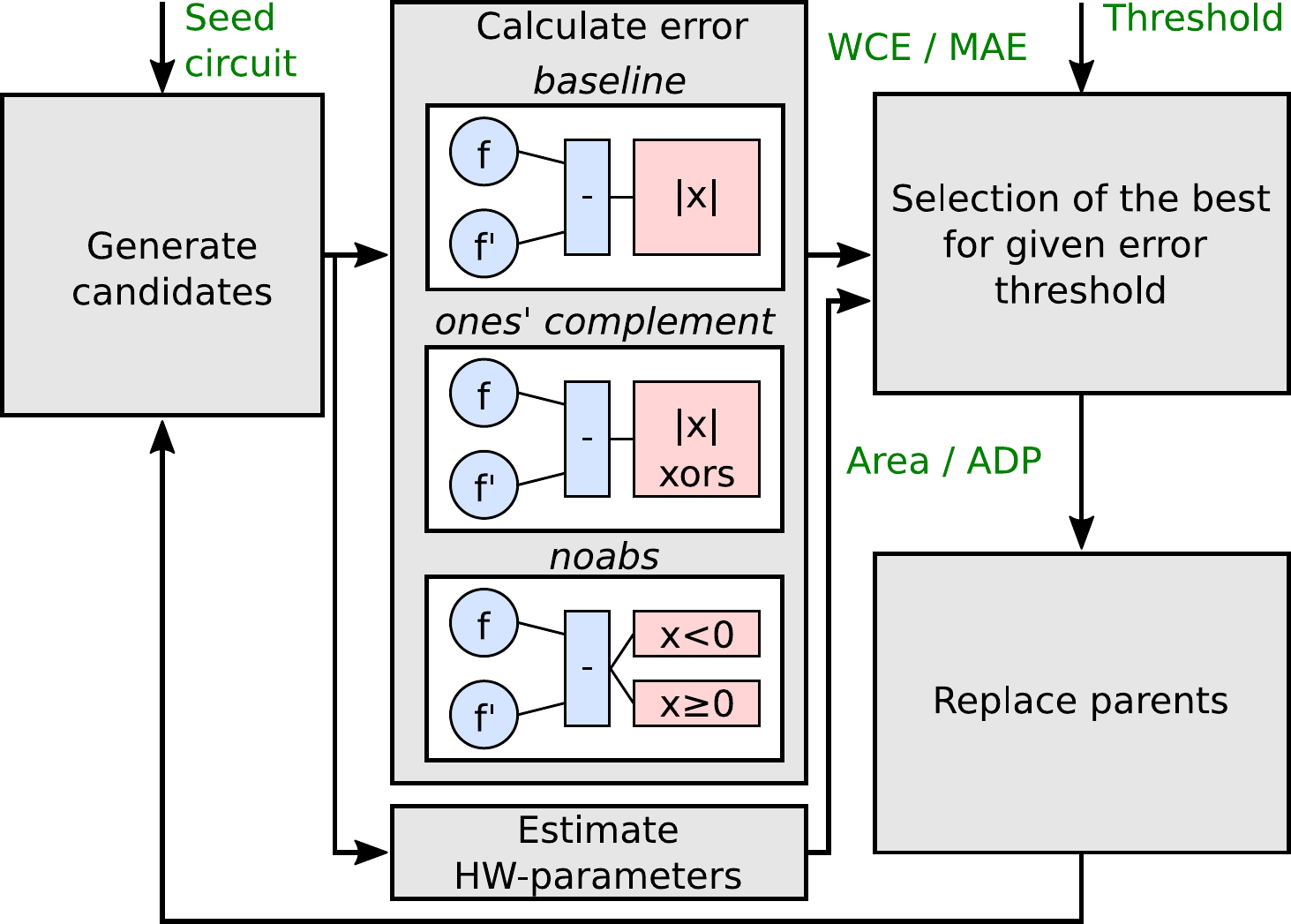}
    \caption{Experimental setup of CGP-based search engine employing the proposed error calculation algorithms.}
    \label{fig:experiments}
\end{figure}

To evaluate the quality of the proposed error calculation algorithms, we employed them in the CGP-based approximation of adders (both signed and unsigned). The overall scheme of the approximation engine is given in Figure \ref{fig:experiments}. In the search loop, tens of thousands of candidate approximate adders were generated and their errors calculated. We collected all results and analyzed them statistically.

The key component of the calculation is the BDD library, which provides an abstraction to BDD trees. It implements the functions deciding the satisfiability $\sat{d}$ and calculating the probability $\satp{d}$. We compared two state-of-the-art libraries --- BuDDy and CuDD. The cache memory of BuDDy was set to $10^7$ accordingly to the documentation. Similarly, default size of CuDD cache ($2^{18} = 262,144$) was used.

	  

The maximal allowed MAE and WCE errors that are the key parameters of CGP exponentially vary from 2 to $0.2 \cdot 2^{n+1}$ (20\% of the output range). As a CGP seed, three adder types are considered: carry-look-ahead, carry-skip, and ripple-carry adders, unsigned and signed versions, for bit-widths 8 to 32 (of each operand). Since the CGP approximation is a stochastic approach, 5 independent runs are executed for each setup. Finally, the best adders are synthesized using Synopsys Design Compiler with 45~nm technology.

All experiments were executed on a computer cluster with nodes having 2 CPU Intel Xeon E5-2670 2,6 GHz, 16-core, 20 MB cache, and 64 GB RAM. Each run was a single thread (with 2 GB of memory), and 32 runs were executed on the node. In total, 49,716 independent runs were performed to evaluate the above-mentioned setups.

\subsection{CGP Approximation with BDD-based error evaluation}
Table~\ref{tab:speedlib} shows that BuDDy library leads to a faster error evaluations than CuDD, except the \textit{baseline} algorithm for 32-bit adders. The results are aggregated over all the error thresholds. 
On average, CuDD is 2x slower than BuDDy. The addition of 8 inputs ($n$) leads to a 4-20x slow down for both libraries.

\begin{table}[ht]
    \centering
    \caption{The average evaluation time [s] for different BDD libraries}
    \label{tab:speedlib}\vspace{-1em}
    \resizebox{\columnwidth}{!}{\begin{tabular}{cc|rr|rr|rr}
    \toprule
    \textbf{Error} & \textbf{Bit-width} & \multicolumn{2}{c|}{\bf Baseline} & \multicolumn{2}{c|}{\bf Ones'} & \multicolumn{2}{c}{\bf Noabs}\\
    &   $n/2$ &\textit{BuDDy} & \textit{CuDD} & \textit{BuDDy} & \textit{CuDD} & \textit{BuDDy} & \textit{CuDD}\\
    \midrule
    
   \multirow{6}{*}{MAE}	&	8	&	\textbf{0.0001}	&	0.0003	&	\textbf{0.0001}	&	0.0002	&	\textbf{0.0001}	&	0.0003\\
	&	12	&	\textbf{0.0009}	&	0.0017	&	\textbf{0.0003}	&	0.0009	&	\textbf{0.0004}	&	0.0013\\
	&	16	&	\textbf{0.0044}	&	0.0084	&	\textbf{0.0013}	&	0.0039	&	\textbf{0.0017}	&	0.0056\\
	&	20	&	\textbf{0.1818}	&	0.2066	&	\textbf{0.0063}	&	0.0223	&	\textbf{0.0113}	&	0.1041\\
	&	24	&	\textbf{1.4713}	&	1.8158	&	\textbf{0.2438}	&	0.6416	&	\textbf{0.3537}	&	0.8289\\
	&	32	&	7.3208	&	\textbf{6.1494}	&	\textbf{3.5416}	&	4.7556	&	\textbf{4.1580}	&	4.7180\\\midrule
\multirow{6}{*}{WCE}	&	8	&	\textbf{0.0001}	&	0.0002	&	\textbf{0.0001}	&	0.0001	&	\textbf{0.0001}	&	0.0001\\
	&	12	&	\textbf{0.0007}	&	0.0010	&	\textbf{0.0002}	&	0.0005	&	\textbf{0.0002}	&	0.0005\\
	&	16	&	\textbf{0.0038}	&	0.0051	&	\textbf{0.0011}	&	0.0023	&	\textbf{0.0009}	&	0.0018\\
	&	20	&	\textbf{0.0446}	&	0.0910	&	\textbf{0.0055}	&	0.0114	&	\textbf{0.0038}	&	0.0072\\
	&	24	&	\textbf{0.8281}	&	0.9924	&	\textbf{0.1044}	&	0.1896	&	\textbf{0.0267}	&	0.0657\\
	&	32	&	5.4822	&	\textbf{4.1673}	&	\textbf{2.2896}	&	2.8677	&	\textbf{2.0623}	&	2.2698\\
    \bottomrule
    \end{tabular}}
\end{table}

Henceforth, only the BuDDy library is used in the following experiments. Table \ref{tab:speedup} shows the average evaluation time for different error evaluation algorithms. Both proposed implementations achieve significant speedup compared to the baseline implementation. On the average, the \textit{ones'} implementation performs better for MAE calculation, but WCE calculation is executed faster without the absolute value (\textit{noabs}). We can see an enormous speedup for 20-bit and 24-bit adders for MAE and WCE, respectively. This is caused by the fact that the proposed algorithms produce smaller BDD trees that can fit the cache. Increasing the cache size could move this peek for larger sizes, but we are also limited by CPU caches and RAM.

\begin{table}[ht]
    \centering
    \caption{An average evaluation time and speedup with BuDDy library}
    \label{tab:speedup}\vspace{-1em}
    \resizebox{\columnwidth}{!}{
    \begin{tabular}{cc|rrr|rr}
    \toprule
    \textbf{Error} & \textbf{Bit-width} & \multicolumn{3}{c|}{\bf Evaluation time [s]} & \multicolumn{2}{c}{\bf Speedup}\\
    &   $n/2$ &\textit{baseline} & \textit{ones'} & \textit{noabs} & \textit{ones'} & \textit{noabs} \\
    \midrule
\multirow{6}{*}{MAE}	&	8	&	0.0001	&	\textbf{0.0001}	&	0.0001	&	\textbf{2.58$\times$}	&	1.91$\times$	\\
	&	12	&	0.0009	&	\textbf{0.0003}	&	0.0004	&	\textbf{3.11$\times$}	&	2.21$\times$	\\
	&	16	&	0.0044	&	\textbf{0.0013}	&	0.0017	&	\textbf{3.47$\times$}	&	2.55$\times$	\\
	&	20	&	0.1818	&	\textbf{0.0063}	&	0.0113	&	\textbf{28.76$\times$}	&	16.06$\times$	\\
	&	24	&	1.4713	&	\textbf{0.2438}	&	0.3537	&	\textbf{6.03$\times$}	&	4.16$\times$	\\
	&	32	&	7.3208	&	\textbf{3.5416}	&	4.1580	&	\textbf{2.07$\times$}	&	1.76$\times$	\\\midrule
\multirow{6}{*}{WCE}	&	8	&	0.0001	&	0.0001	&	\textbf{0.0001}	&	2.37$\times$	&	\textbf{2.48$\times$}	\\
	&	12	&	0.0007	&	0.0002	&	\textbf{0.0002}	&	2.84$\times$	&	\textbf{3.19$\times$}	\\
	&	16	&	0.0038	&	0.0011	&	\textbf{0.0009}	&	3.33$\times$	&	\textbf{4.20$\times$}	\\
	&	20	&	0.0446	&	0.0055	&	\textbf{0.0038}	&	8.13$\times$	&	\textbf{11.89$\times$}	\\
	&	24	&	0.8281	&	0.1044	&	\textbf{0.0267}	&	7.93$\times$	&	\textbf{31.04$\times$}	\\
	&	32	&	5.4822	&	2.2896	&	\textbf{2.0623}	&	2.39$\times$	&	\textbf{2.66$\times$}	\\
	\bottomrule
    \end{tabular}}
\end{table}

We separately analyzed the impact of signed and unsigned versions of adders. The signed adders need the subtractor in the characteristic function to be extended by one bit. This extension slows down the error calculation by 3 - 17\% for the proposed algorithms.

\subsection{Error characterisation of 16-bit adders}
\begin{figure}[ht]
    \centering
    \vspace*{-1em}
    \includegraphics[width=\columnwidth]{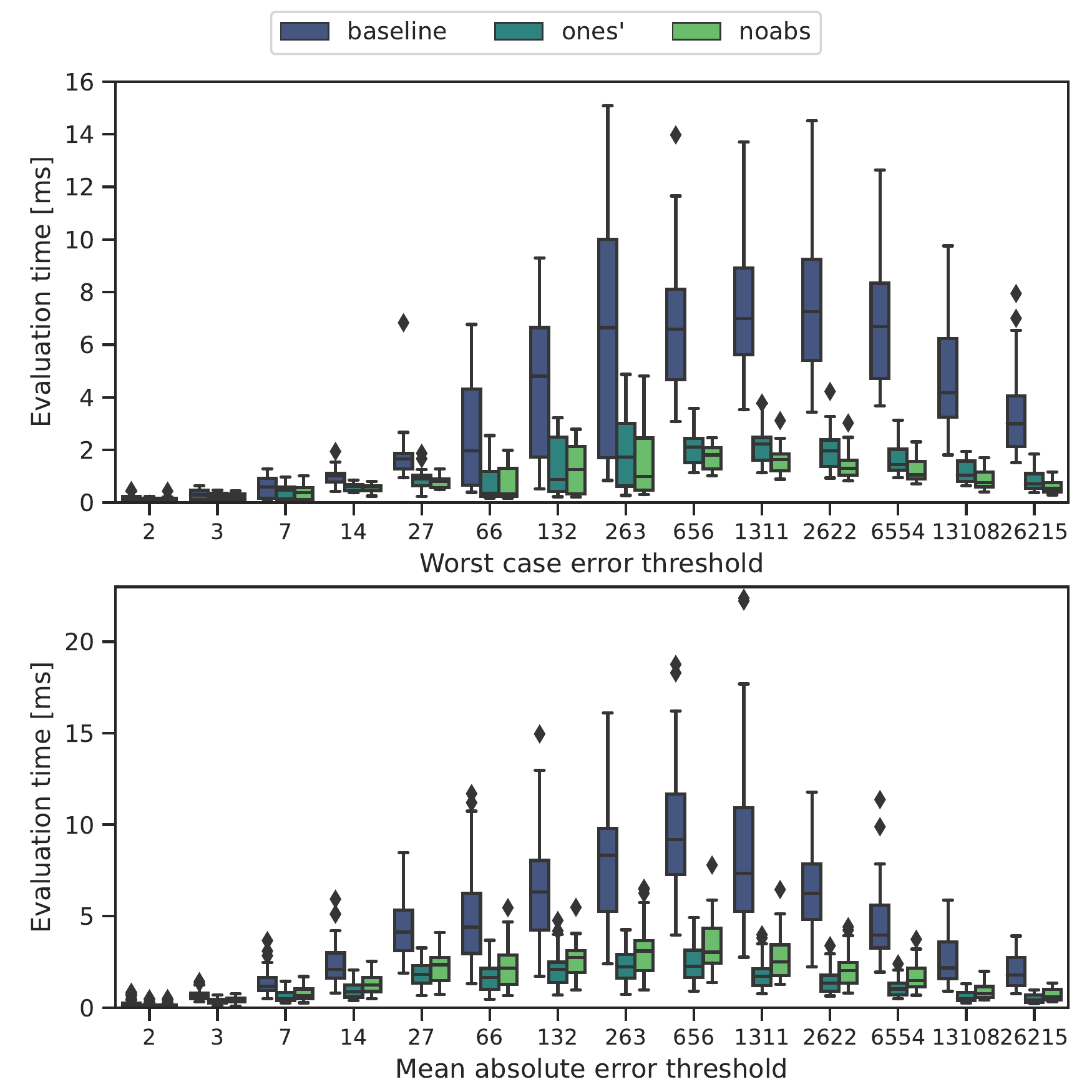}
    \vspace*{-2em}
    \caption{Effect of the threshold value $\tau$ on the average evaluation time in CGP-based approximations.}
    \label{fig:speederror}
\end{figure}

We analyzed the behavior of the proposed error calculation algorithms in the design of  16-bit approximate adders in greater detail. 
The average evaluation time depends on the selected error analysis algorithm,  error threshold, and the seed circuit. Figure \ref{fig:speederror} shows the average evaluation time for different error thresholds. The slowest evaluations are in the middle part of thresholds regardless of the evaluation algorithm. It is caused by the following: (i) For small thresholds, the BDDs representing $f'$s  are not much different from the golden circuits $f$.(ii) For large thresholds, BDDs representing $f'$s are simplified (because of the approximation), and their sizes are reduced. We can see the proposed algorithms achieved significantly faster evaluations than the baseline.

One of the inputs of the approximation algorithm is the seed circuit. This circuit acts as a parent while all candidate solutions are derived using a chain of mutations. We use this circuit as a golden circuit $r$ in the characteristic function $\varepsilon = f - f'$. Interestingly, the speed is not related to the size of the circuit (CLA:171, CSkA: 129, and RCA: 77 nodes, signed versions have two additional nodes). Although the RCA is the smallest one (and produces more approximated circuits), the evaluation speed is lower.

\begin{figure}[ht]
    \centering
    \includegraphics[width=\columnwidth]{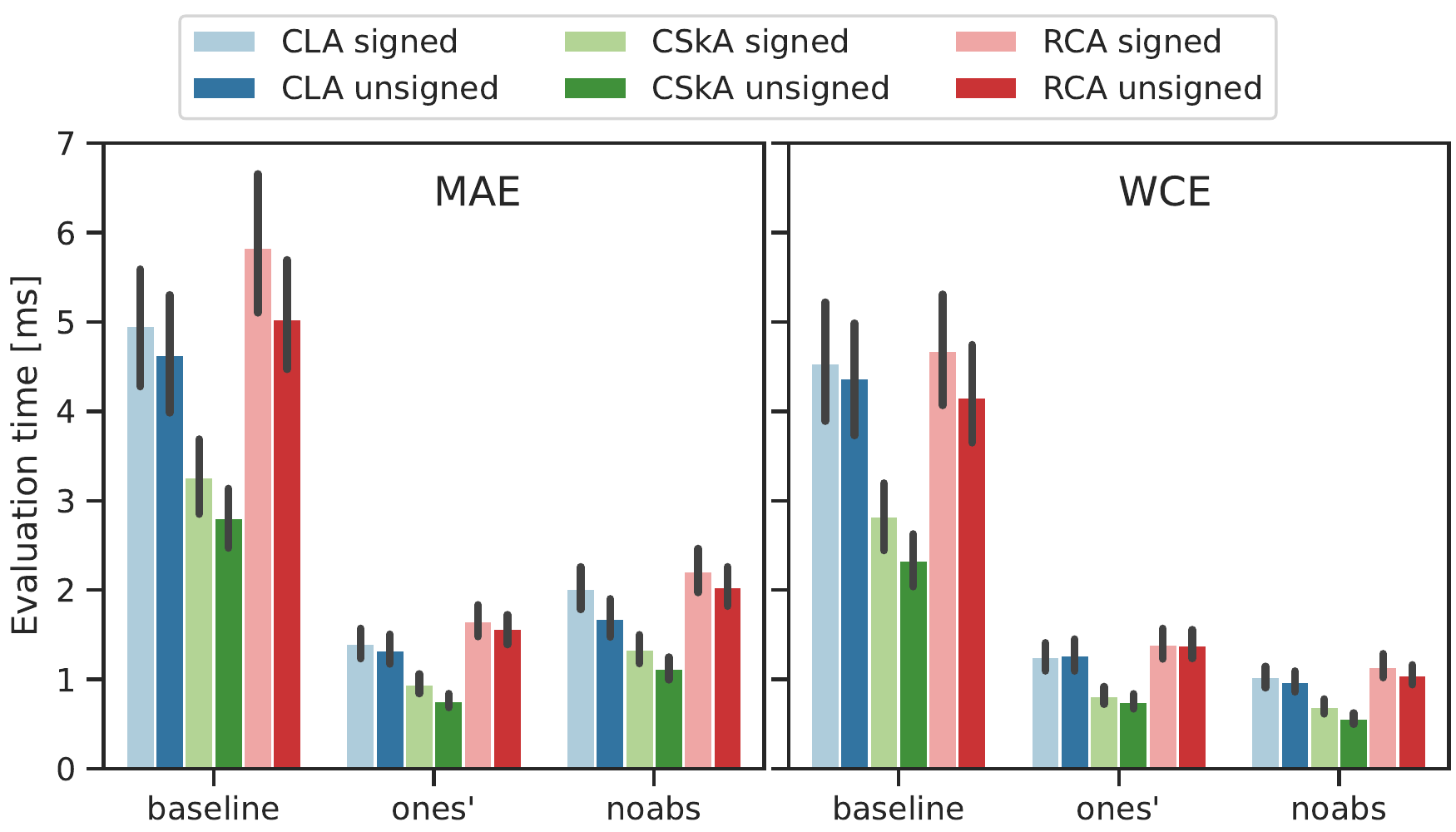}
    \vspace*{-2em}
    \caption{Effect of selected seed and golden model implementation for the average evaluation speed}
    \vspace*{-1em}
    \label{fig:seeds}
\end{figure}

\begin{figure}[ht]
    \centering
    \includegraphics[width=\columnwidth]{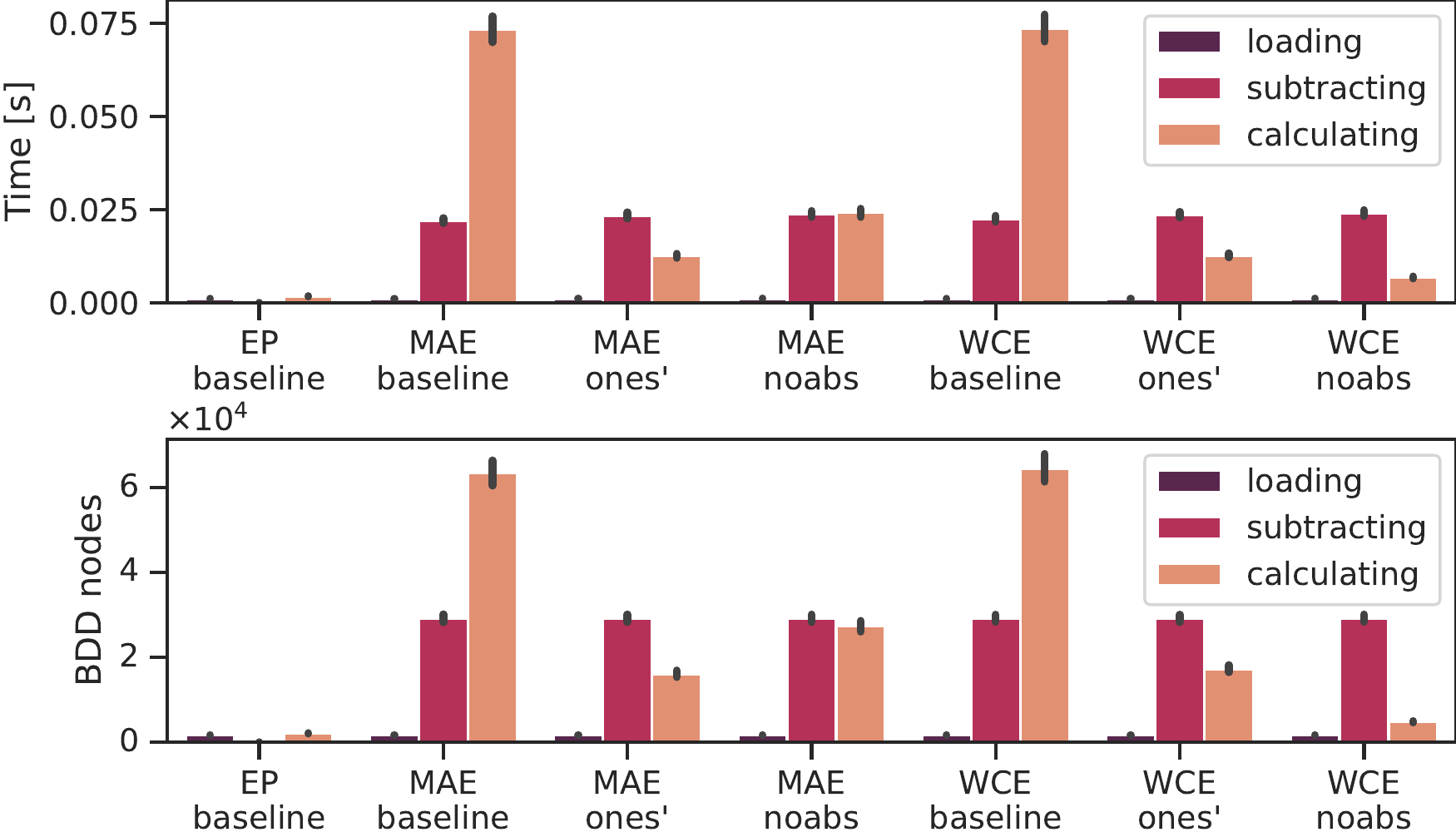}
    \vspace*{-2em}
    \caption{Time spent and number of created nodes in three steps of the error calculation algorithms. Error probability $e_{ep}$~\cite{soeken:bdd} is included for complexity illustration.}
    \vspace*{-1em}
    \label{fig:parts}
\end{figure}

We analyzed the error-metric evaluation for all resulting approximate implementations of 16-bit adders using considered BDD-based algorithms. Figure \ref{fig:parts} illustrates the execution time and the total number of created BDD nodes for three main parts of the error evaluation procedure. The part named \textit{loading} represents transforming the netlist of functions $f$ and $f'$ into BDDs. The second part, called \textit{subtracting}, consists of the ripple-carry subtractor creation. The last part, called \textit{calculating}, includes building BDDs that represent the calculation of the absolute value and the final evaluation. We can see that the first part is negligible in contrast to the subtracting and calculating parts. The \textit{baseline} implementation has an enormous proportion of the \textit{calculating} part, but the number of BDD nodes is relatively smaller. However, the overall evaluation time roughly correlates with the number of BDD nodes.

\subsection{Synthesis and comparison to SoA}
\begin{figure}[ht]
    \centering\vspace*{-1em}
    \includegraphics[width=\columnwidth]{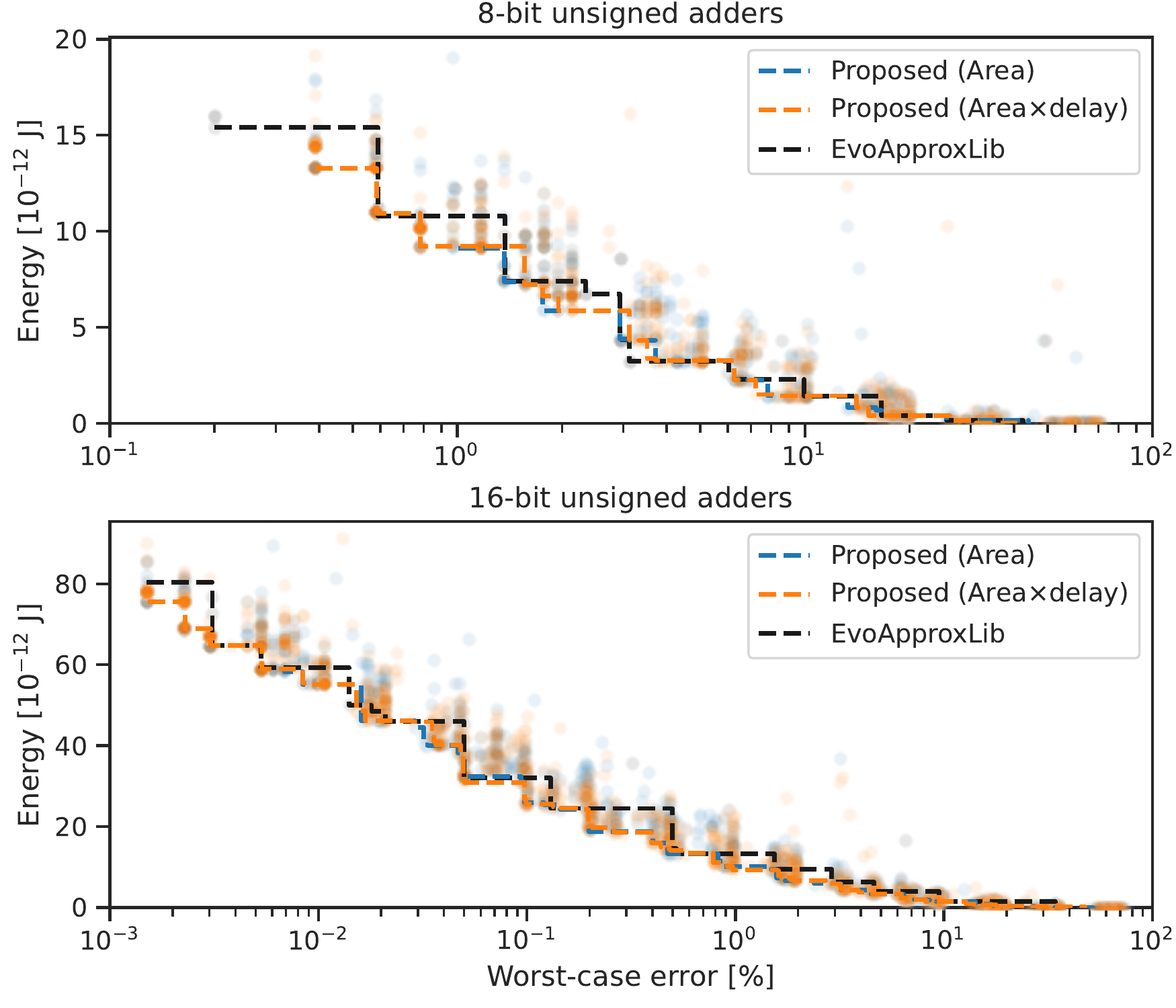}
    \vspace*{-2em}
    \caption{Comparison of synthesized circuits designed with CGP with two fitness functions using proposed error calculations.}
    \vspace*{-1em}
    \label{fig:synthwce}
\end{figure}

The resulting approximate adders are synthesized to obtain hardware parameters (we focus on energy). Both error metrics are evaluated. Figure \ref{fig:synthwce} shows trade-offs between $e_{wce}$ and the energy. 
In addition to the area estimation in the fitness function (eq. \ref{eq:fitness}), we also used CGP with the area-delay product (ADP) in the fitness function. However, this setup did not improve the results in terms of energy. 
The resulting approximate adders outperform relevant approximate adders published in the EvoApproxLib library~\cite{mrazek:date16lib}). Moreover, we designed larger approximate adders, including the signed and unsigned variants that are not included in the library. These circuits were introduced to the new version of EvoApproxLib.



\section{Conclusion}
We proposed new BDD-based algorithms for the fast error evaluation of approximate adders and subtractors. Compared with the baseline implementation, these algorithms reduced the evaluation time 3x times on average. Moreover, they do not need any assumption about $f'$. These algorithms enabled us to accelerate the approximate circuit design process. Our future work will be devoted to adapting the principle for different arithmetic error metrics and the application of the proposed circuits.

\vspace{1mm}
{\textit{Acknowledgement} This work has been supported the Czech Science Foundation grant (GJ20-02328Y).}


\bibliographystyle{./bibliography/IEEEtran}
\bibliography{./bibliography/IEEEabrv,bdd_add}

\end{document}